\documentclass[10pt,journal,final]{IEEEtran}
\ifCLASSOPTIONcompsoc
   \usepackage[nocompress]{cite}
\else
   \usepackage{cite}
\fi
\usepackage{color}
\usepackage{graphicx}
\usepackage{psfrag}
\usepackage{lscape}
\usepackage{amssymb}
\usepackage{amsmath}
\usepackage{cases}
\usepackage{algorithm}
\usepackage{algorithmic}
\ifCLASSOPTIONcompsoc
\usepackage[tight,normalsize,sf,SF]{subfigure}
\else
\usepackage[tight,footnotesize]{subfigure}
\fi

\begin{document}
%
\title{Dual Decomposition-Based Privacy-Preserving Multi-Horizon Utility-Community Decision Making Paradigms}
\author{Vahid R. Disfani,~\IEEEmembership{Student Member,~IEEE,}
        Zhixin Miao, ~\IEEEmembership{Senior Member,~IEEE,}
        Lingling Fan,~\IEEEmembership{Senior Member,~IEEE,}
        Bo Zeng,~\IEEEmembership{Member,~IEEE}
\thanks{V.R. Disfani, Z. Miao, and L. Fan are with the Department of Electrical Engineering, University of South Florida, Tampa, FL 33620 (Email: linglingfan@usf.edu). B. Zeng is with the Department of Industrial Management and Science at University of South Florida.}
}

\maketitle
\begin{abstract}

Two types of privacy-preserving decision making paradigms for utility-community interactions for multi-horizon operation are examined in this paper. In both designs, communities with renewable energy sources, distributed generators, and energy storage systems minimize their costs with limited information exchange with the utility. The utility makes decision based on the information provided from the communities. Through an iterative process, all parties achieve agreement. The authors' previous research results on subgradient and lower-upper-bound switching (LUBS)-based distributed optimization oriented  multi-agent control strategies are examined and the convergence analysis of both strategies are provided. The corresponding decision making architectures, including information flow among agents and learning (or iteration) procedure, are developed for multi-horizon decision making scenarios. Numerical results illustrate the decision making procedures and demonstrate their feasibility of practical implementation. The two decision making architectures are compared for their implementation requirements as well as performance.

\end{abstract}
\begin{IEEEkeywords}
AC OPF, dual decomposition, moving horizon optimization, convergence property, spin reserve 
\end{IEEEkeywords}

\section{Introduction}
\IEEEPARstart{S}IGNIFICANT increase in penetration of private agents having their own microgrids to the power network is highly expected in the future smart grid, which makes control and optimization of the power network more challenging than before. These microgrids may have different types of distributed energy resources (DER) such as renewable energy systems (RES) and energy storage systems (ESS) to mitigate renewable energy intermittency. Storage systems are constrained by temporal limitations which makes decision making to consider multiple time horizons instead of just one snapshot. In addition, the private agents have their own specific interests and do not like to share their private information with the system operator due to economical or privacy-related reasons. From the operator's point of view, it is desired to operate the entire power network including the microgrids in its optimal operation. In such optimization problems, the network constraints such as network congestion and voltage limits must be considered. Furthermore, the spinning reserve is required, which guarantees the system be ready to respond to contingencies that affect power balance.

The \textbf{\emph{objective}} of this paper is to develop privacy-preserving multi-horizon utility-community decision making paradigms considering AC power flow, network constraints, and reserve requirements.

The mathematic foundation of privacy-preserving decision making is distributed optimization. In the literature, there are several approaches to develop distributed algorithms for both AC OPF \cite{kim1997coarse,kim2000comparison,erseghe2012distributed,dall2013distributed,disfani2014multi} and DC OPF \cite{conejo1998multi,bakirtzis2003decentralized,biskas2005decentralized,disfani2015distributed}. These algorithms decompose an optimization problem to smaller subproblems and design an iterative process to seek the optimal solution, where one agent is responsible to each one of these subproblems. After the agents solve their own OPF problem (e.g, using using interior point method), these algorithms performs an update process to push the solution toward the optimal solution. The update process requires agents to share information. Therefore, each algorithm corresponds to an information flow structure.


A modified subgradient-based method has been proposed in the authors' previous work \cite{disfani2014multi}, where microgrids and a utility name their desired levels of power transactions and a price updating scheme pushes the solution toward the optimal solution. Another method proposed in \cite{disfani2014multi} is named lower-upper-bound-witching (LUBS), where the microgrids update their desired prices based on the power demand of the main grid. The main grid defines its optimal power export/import level from each microgrid based on the prices. The convergence of the upper-bound and the lower-bound indicates the optimality of the solution.

To handle the time-correlated constraints of ESS, OPF need to consider multiple time horizons instead of just one snapshot. Several approaches has been addressed in the literature to tackle multi-horizon OPF problem. Bender's decomposition method is developed in \cite{alguacil2000multiperiod} to tackle energy constraints of hydroelectric plants integrated in irrigation systems. Bender' decomposition focuses on decomposition of integer decision variables and continuous decision variables and cannot be translated into a multi-agent control structure. Another thread of research \cite{chandy2010simple,gayme2013optimal} develops a KKT-based solution to consider battery storage systems. An optimization model is also proposed in \cite{gabash2012active} for multi-horizon OPF with wind generation and battery storage. The capability of ESS to provide ancillary services such as reserve is taken into account to tackle reserve constraints in an OPF problem without considering their time-correlated constraints \cite{sortomme2009optimal,wen2014enhanced}.

To the authors' best knowledge, the proposed research privacy-preserving decision making for multi-horizons while considering AC OPF and spinning reserve, has not been seen in the literature. This paper extends the research in \cite{disfani2014multi} to apply the privacy-preserving decision making to multi-horizons while considering spinning reserve requirements. Spinning reserve adequacy is a constraint which is considered not only in unit commitment problem \cite{shahidehpour2002frontmatter} but also in OPF problem \cite{condren2006optimal,zimmerman2009matpower}. A reserve-constrained OPF problem is developed in \cite{condren2006optimal} to incorporate the expected security costs in the system using reserve marginal value (reserve price) which indicates how much cost is imposed to the system operation if one more unit of reserve is needed. The extension of MATPOWER 4.1 \cite{matpower} can solve OPF with co-optimization of energy and reserve. This capability is employed in our research to solve reserve-constrained OPF problems.

The contribution of this paper has threefold.
\begin{itemize}
\item Two privacy-preserving decision making paradigms based on our previous research are developed for multi-horizons while considering reserve constraints;
\item Convergence properties for the paradigms are analyzed and convergence enhancement measures are proposed and explained.
\item Realistic implementation of the decision making paradigms at real time is investigated using moving horizon optimization technique and demonstrated in case studies.
\end{itemize}


The rest of the paper is organize as following: the two dual decomposition-based paradigms for a multi-horizon OPF problem are presented in Section II. Implementation for real-time operation is investigated and moving horizon optimization technique is adopted. The algorithms' convergence properties are investigated in Section III.  Section VV presents a case study to demonstrate the decision making process. The conclusion is presented in Section V.
%
%
%
%

\section{Dual Decomposition-Based Privacy-Preserving Decision Making Paradigms}
\begin{figure}[htbp]
	\centering
	\includegraphics[width=0.20\textwidth]{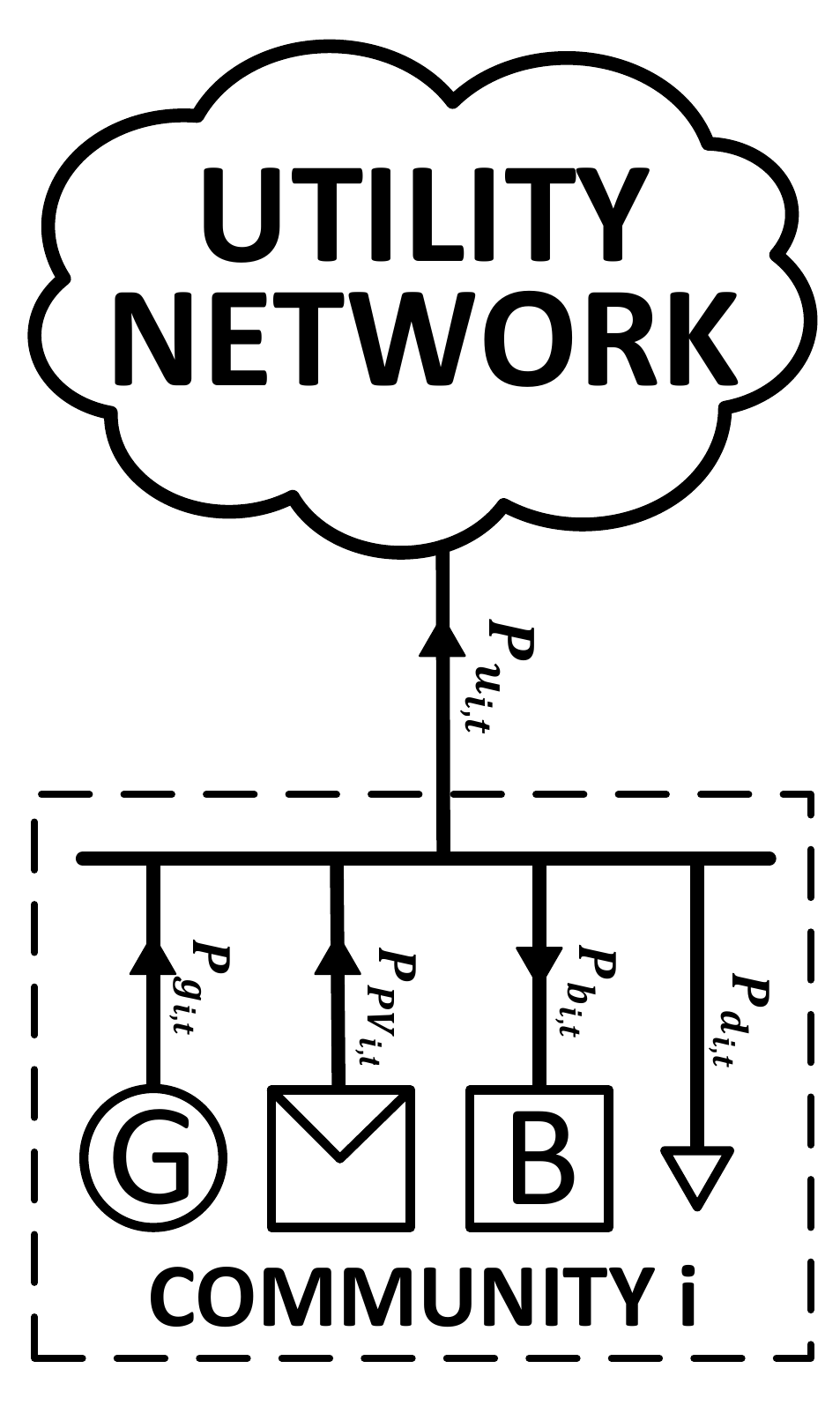}
	\caption{A community microgrid including a generator, a renewable energy system (RES), an energy storage system (ESS) and internal load.}
	\label{standalone_fig}
\end{figure}

Consider a power network consisting of a set $\mathcal{N}$ of buses and a set $\mathcal{E}$ of branches. The utility is responsible to operate the power grid, its generation units and transactions with transmission systems. As shown in Fig. \ref{standalone_fig}, community microgrids may have energy storage systems, renewable energy resources, and internal loads. These microgrids are connected to the network and behave as private agents who will share only limited information with the utility. The set $\mathcal{A}$ includes all buses which the communities are connected to.

The reserve-constrained 24-hour OPF problem considering community microgrids is defined as below:
     \begin{subequations}
     \begin{align}
        \min \>\>\>\>\>& \sum_{t=1}^{24}\sum_{i\in \mathcal{N}} C_i(P_{g_i})\label{MH_OPF_obj}\\
\text{s. t.} \>\>\>\>\>& \forall_{i\in \mathcal{N}},\>\forall_{j\in \mathcal{E}},\>\forall_{t\in \mathcal{T}}\nonumber\\
     & P_{g_{i,t}}-P_{L_{i,t}}+P_{PV_{i,t}}-P_{b_{i,t}}-P_{i,t}(V, \theta) =0  \label{MH_OPF_active_balance}\\
     & Q_{g_{i,t}}-Q_{L_{i,t}}+Q_{PV_{i,t}}-Q_{b_{i,t}}-Q_{i,t}(V, \theta) =0  \label{MH_OPF_reactive_balance}\\
     & E_{b_{i,t+1}}= E_{b_{i,t}}+ P_{b_{i,t}} \label{MH_OPF_battery_enrg_power}\\
     & V_i^m \le V_{i,t} \le V_i^M \label{MH_OPF_volt_limit}\\
     & P^m_{g_i} \le P_{g_{i,t}} \le P^M_{g_i} \label{MH_OPF_active_gen_limit}\\
     & Q^m_{g_i} \le Q_{g_{i,t}} \le Q^M_{g_i}  \label{MH_OPF_reactive_gen_limit}\\
     & S_{j,t}(V, \theta)-S_j^M \le 0 \label{MH_OPF_line_power_limit}\\
     & P^m_{b_i} \le P_{b_{i,t}} \le P^M_{b_i} \label{MH_OPF_battery_power_limit}\\
     & E^m_{b_i} \le E_{b_{i,t+1}} \le E^M_{b_i} \label{MH_OPF_battery_enrg_limit}\\
     & E_{b_{i,24}}= E_{b_{i,0}} \label{MH_OPF_battery_same_enrg}\\
     & \sum_{i}{R_{g_{i,t}}+R_{b_{i,t}}}\ge R_{d_t}\label{MH_OPF_reserve_satisfy}\\
     & R_{g_{i,t}}\le R_{g_{i}}^M\label{MH_OPF_reserve_const1}\\
     & R_{g_{i,t}}\le P_{g_{i}}^M-P_{g_{i,t}}\label{MH_OPF_reserve_const2}\\
     & R_{b_{i,t}}\le -P_{b_{i}}^m+P_{b_{i,t}}\label{MH_OPF_reserve_battery}
      \end{align}
     \label{MH_OPF}
     \end{subequations}
     where $C(\cdot)$ is the cost function, superscripts $M$ and $m$ denote upper and low limits.
     Subscripts $i\in\mathcal{N}$ and $t\in\mathcal{T}$ refer to the variables corresponding to bus $i$ and hour $t$ where the set $\mathcal{T}=\{t\in\mathbb{Z}|1\le t\le24\}$ denotes the time horizon.
     $P_g$, $Q_g$, $P_L$ and $Q_L$ are the vectors of bus real and reactive power generations, and real and reactive loads.
     $P_b$ and $P_{PV}$ denote the battery charging power and PV output power respectively.
     $P(V, \theta)$ and $Q(V, \theta)$ are the active and reactive power injections in terms of bus voltage magnitude and phase angles,
     and $S(V, \theta)$ is the vector of line complex power flow.
     $E_{b}$ also determines the energy stored in the storage systems.
     The parameters $R_g$ and $R_b$ denote the reserve provided by the generators and battery systems respectively.

     In the optimization problem \eqref{MH_OPF}, the objective function \eqref{MH_OPF_obj} is defined so as to minimize the total generation cost in the entire system. Active and reactive power balance constraints are defined in \eqref{MH_OPF_active_balance} and \eqref{MH_OPF_reactive_balance} for all buses at all hours. The dependency between battery charging power and energy stored in the battery is stated in \eqref{MH_OPF_battery_enrg_power}. The constraints \eqref{MH_OPF_volt_limit}-\eqref{MH_OPF_battery_enrg_power} describe minimum and maximum limits of the corresponding variables. The constraint \eqref{MH_OPF_battery_same_enrg} mandates that the ultimate level of energy stored in the battery to be equal to that at starting point. The minimum level of total reserve from both utility generators and the community energy sources is guaranteed by \eqref{MH_OPF_reserve_satisfy} while the constraints \eqref{MH_OPF_reserve_const1}-\eqref{MH_OPF_reserve_battery} take care of the reserve constraints.

Let $P_{\rm imp_{i,t}}$ denotes the utility's power import from community connected to bus $i\in\mathcal{A}$ at time $t$ while $P_{\rm exp_{i,t}}$ denotes the same community's power export to the utility. Therefore, the power balance equations at Bus $i$ will be replaced by the following four equations.
     \begin{align}
    & P_{\rm imp_{i,t}}=P_{i,t}(V, \theta) \notag\\
    & P_{\rm exp_{i,t}}=P_{g_{i,t}}-P_{L_{i,t}}+P_{PV_{i,t}}-P_{b_{i,t}} \notag\\
    & Q_{\rm imp_{i,t}}=Q_{i,t}(V, \theta) \notag\\
    & Q_{\rm exp_{i,t}}=Q_{g_{i,t}}-Q_{L_{i,t}}+Q_{PV_{i,t}}-Q_{b_{i,t}}
     \label{expimp}
     \end{align}

A community now only relates to $P_{\rm exp_{i,t}}$, not $P_{\rm imp_{i,t}}$. Global constraints that relate the communities to the utility are imposed as 
\begin{align}
     \begin{aligned}
      \lambda^p_{i,t}:\>\> & P_{\rm imp_{i,t}}=P_{\rm exp_{i,t}},\\
      \lambda^q_{i,t}:\>\> & Q_{\rm imp_{i,t}}=Q_{\rm exp_{i,t}},\\
      \mu_{t}:\>\>& R_{d_t}-\sum_{i\in\mathcal{N-A} }{R_{g_{i,t}}-\sum_{j\in \mathcal{A}}R_{{j,t}}}\le 0 .
          \label{expimp}
     \end{aligned}
     \end{align}
     with the corresponding Lagrangian multipliers also notated. In the reserve constraint, we have separated reserve provided by the utility generators $R_{g_{i,t}}$ and the reserve provided by a community $j$ $R_{j,t}$. Note there are multiple real power and reactive power prices at $t$. However there is one reserve price at $t$.

Applying Lagrangian relaxation to relax the three global constraints in \eqref{expimp}, the main problem is divided into several subproblems each of which is solved by the utility or one of the communities. The subproblem assigned to the utility is:
     \begin{align}
        \min \>&  \sum_{t=1}^{24}{\sum_{i\in\mathcal{N-A}} [C_i-\mu_{t}{R_{g_{i,t}}}]+ \sum_{j\in\mathcal{A}}[\lambda^p_{j,t}P_{{\rm imp}_{j,t}}+\lambda^q_{j,t}Q_{{\rm imp}_{j,t}}]} \label{MH_SG_Utility}\\
        \text{over} \>\> & P_{g_{i,t}}, Q_{g_{i,t}}, P_{\rm imp_{i,t}}, Q_{\rm imp_{i,t}}, R_{g_{i,t}}\\
        \text{s. t.} \>\> & P^m_{g_{i,t}} \le P_{g_{i,t}} \le P^M_{g_{i,t}} \nonumber\\
            & Q^m_{g_{i,t}} \le Q_{g_{i,t}} \le Q^M_{g_{i,t}}  \nonumber\\
             & R_{{i,t}}\le R_{g_{i}}^M,\>\> R_{{i,t}}\le P_{g_{i}}^M-P_{g_{i,t}}\\
             & R_{{i,t}}\ge0, \>\> R_{\rm imp_{i,t}}\ge0.\nonumber
         \end{align}

The community $i$ solves the following problem at each hour $t$.
     \begin{align}
     \begin{aligned}
        \min \>\>\>& \sum_{t=1}^{24} [C_i(P_{g_{i,t}})-\lambda_{i,t}^p{P_{\rm exp_{i,t}}}-\lambda_{i,t}^q{Q_{\rm exp_{i,t}}}-\mu_{t}{R_{{i,t}}}] \\
        \text{over} \>\>\>&{P_{g_{i,t}},Q_{g_{i,t}},P_{b_{i,t}}}\\
        \text{s. t.} \>\>\>   & P_{\rm exp_{i,t}}=P_{g_{i,t}}-P_{L_{i,t}}+P_{PV_{i,t}}-P_{b_{i,t}}\\
             & Q_{\rm exp_{i,t}}=Q_{g_{i,t}}-Q_{L_{i,t}}+Q_{PV_{i,t}}-Q_{b_{i,t}}\\
             & P^m_{g_{i,t}} \le P_{g_{i,t}} \le P^M_{g_{i,t}} \\
             & Q^m_{g_{i,t}} \le Q_{g_{i,t}} \le Q^M_{g_{i,t}} \\
            & Q^m_{PV_i} \le Q_{PV_{i,t}} \le Q^M_{PV_i} \\
            & P^m_{b_i} \le P_{b_{i,t}} \le P^M_{b_i} \\
            & Q^m_{b_i} \le Q_{b_{i,t}} \le Q^M_{b_i} \\
             & E_{b_{i,t+1}}= E_{b_{i,t}}+ P_{g_{i,t}}\\
             & E^m_{b_i} \le E_{b_{i,t+1}} \le E^M_{b_i}\\
             & E_{b_{i,25}}= E_{b_{i,1}} \\
             & R_{{i,t}}={R_{g_{i,t}}}+{R_{b_{i,t}}}\\
             & R_{g_{i,t}}\le R_{g_{i}}^M\\
            & R_{g_{i,t}}\le P_{g_{i}}^M-P_{g_{i,t}}
             \end{aligned}
     \label{MH_SG_Comm}
     \end{align}

\begin{figure}[htbp]
	\centering
	\includegraphics[width=0.45\textwidth]{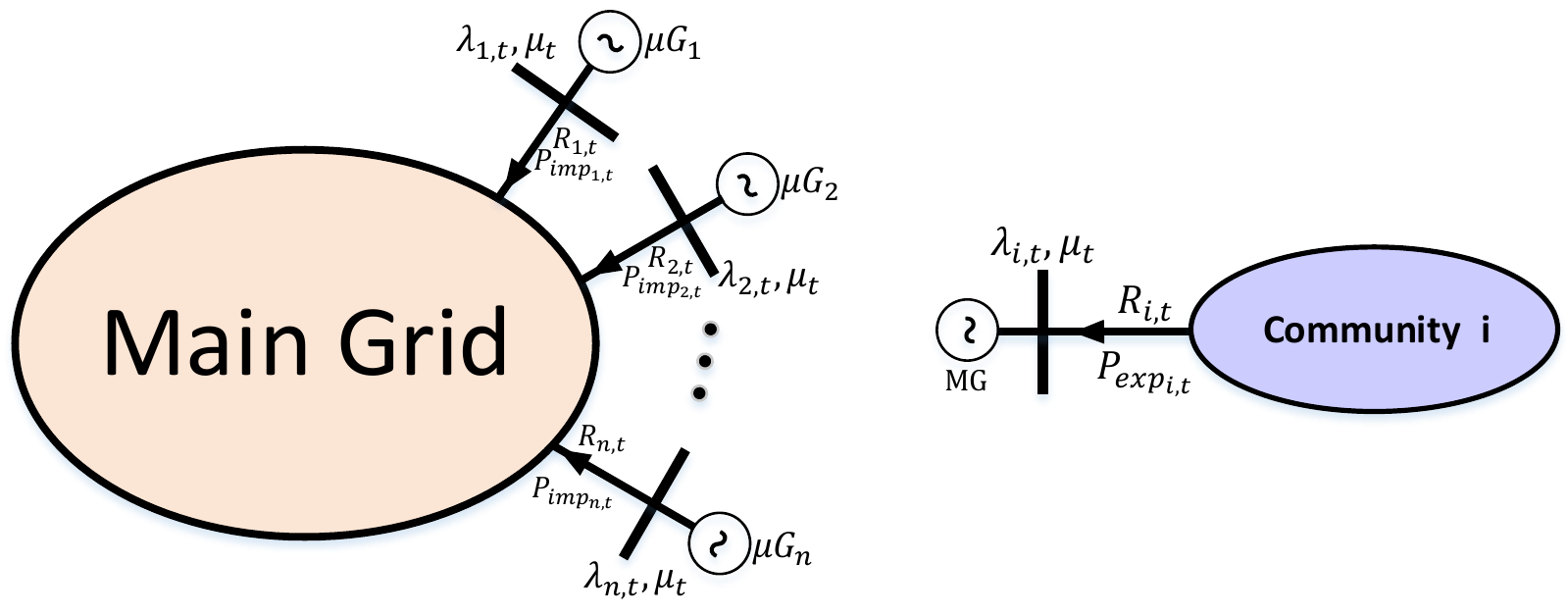}
	\caption{Power networks in utility and community optimization problems.}
	\label{fig:utilcomgen}
\end{figure}

Fig. \ref{fig:utilcomgen} gives an illustration to show the decomposed systems where the utility will treat every community as a generator with prices of energy and reserve given and the community will treat the main grid as a controllable load with prices of energy and reserve also given.

\subsection{Subgradient-Based Method}
\begin{figure}[htbp]
	\centering
	\includegraphics[width=0.45\textwidth]{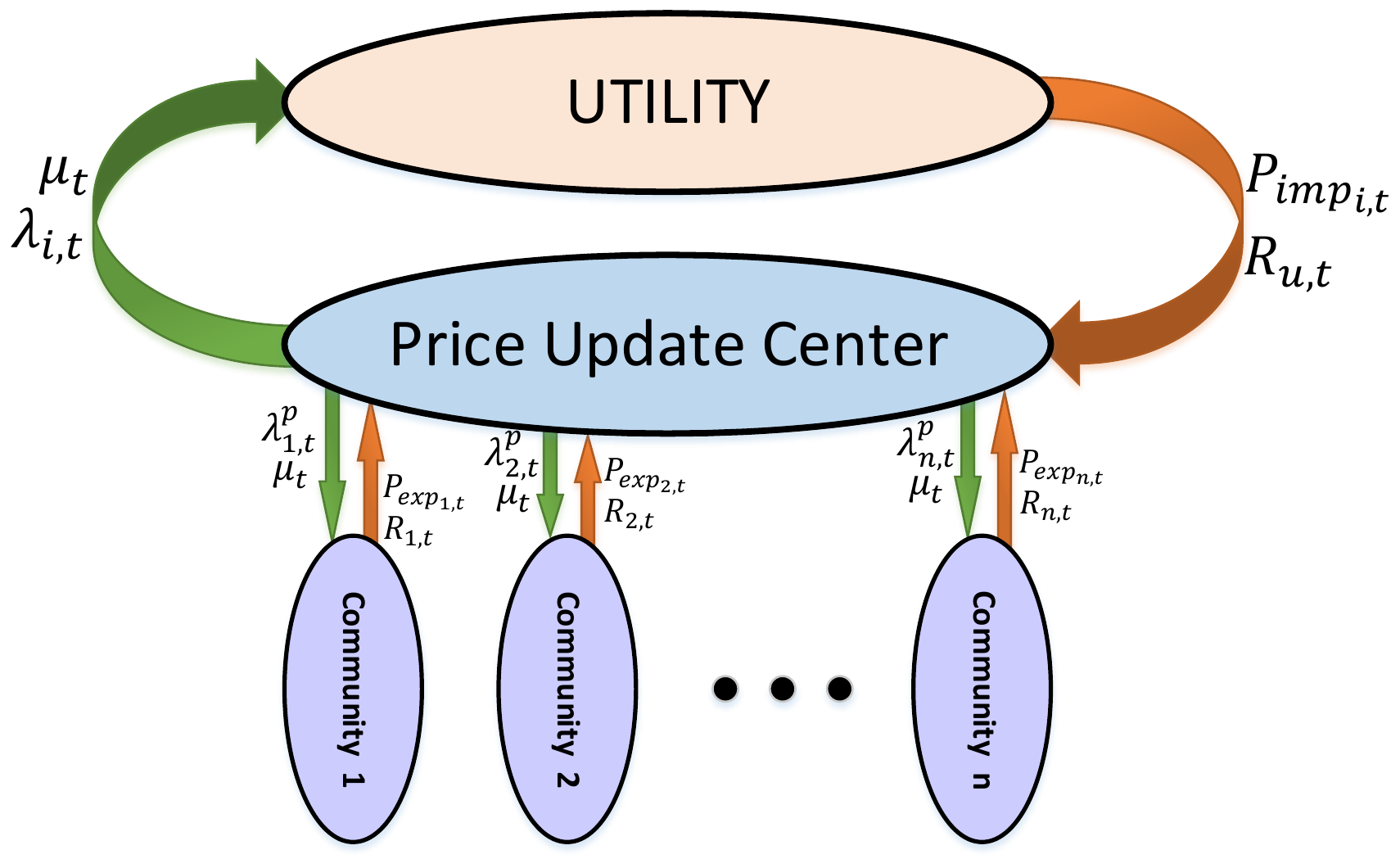}
	\caption{Information flow of multi-horizon subgradient algorithm. In this figure, information related to reactive power is not notated.}
	\label{fig:SGinfoflow}
\end{figure}

In subgradient method, the energy price and reserve price signals are first specified by the price update center (PUC) as illustrated in Fig. \ref{fig:SGinfoflow}. The price signals are used by the utility and communities to define their power import and export levels as well as reserve for all $t\in\mathcal{T}$. For the next iteration, the energy price values $\lambda_{i,t}$ for the next iteration are then updated using the mismatch between corresponding power import and export levels. The reserve price value $\mu_{t}$ is also updated to reflect the mismatch of import and export level. Fig. \ref{fig:SGinfoflow} also depicts the information flow between these three blocks through the algorithm.
\subsubsection{Price Updating}
Updating the energy prices at each bus and the reserve price are the functions assigned to Price Update Center, utilizing the following equations:
      \begin{align}
        &\lambda^p_{i,t}(k+1)=\lambda^p_{i,t}(k)+\alpha_{i,t}(k) \cdot (P_{\rm imp_{i,t}}-P_{\rm exp_{i,t}})\\
        &\lambda^q_{i,t}(k+1)=\lambda^q_{i,t}(k)+\alpha_{i,t}(k) \cdot (Q_{\rm imp_{i,t}}-Q_{\rm exp_{i,t}})\\
        &\mu_{t}(k+1)=\mu_{t}(k)+\beta_{t}(k) \cdot ( R_{d_t}-\sum_{i\in\mathcal{N-A} }{R_{g_{i,t}}-\sum_{j\in \mathcal{A}}R_{{j,t}}})
     \label{EQ:conv_SG}
     \end{align}
where $k$ is the index of updating step, $\alpha_{i,t}(k)$ and $\beta_t(k)$ are positive coefficients.

\subsubsection{Community Optimization}
In order to define its power export level at different hours, each community must perform its optimization based on the given price from the PUC. In this paper, it is assumed that each community has a generator and a defined power load as well as a PV-battery package. The optimization problem corresponding to the community connected to bus $i\in \mathcal{A}$ is presented in \eqref{MH_SG_Comm}.

Each community receives the updated price for energy and reserve, solves its own subproblem \eqref{MH_SG_Comm} and determines the amount of power to sell or to export to the main grid ($P_{\rm exp_{i,t}}$ for all $t\in\mathcal{T}$) as well as the level of reserve it may provide. 

\subsubsection{Utility Optimization}
For each hour $t$, the utility updates the import level. In this paper, MATPOWER 4.1 \cite{matpower} is employed to solve the reserve-constrained OPF problem where the the communities are assumed as traders selling their power by the price announced by PUC. These values create the power import matrix $P_\text{imp}$ from different communities at different hours as well as the reserve vector $R$, which are sent to the PUC for price updating process for the next iteration.

\subsection{LUBS Method}
\begin{figure}[htbp]
	\centering
	\includegraphics[width=0.45\textwidth]{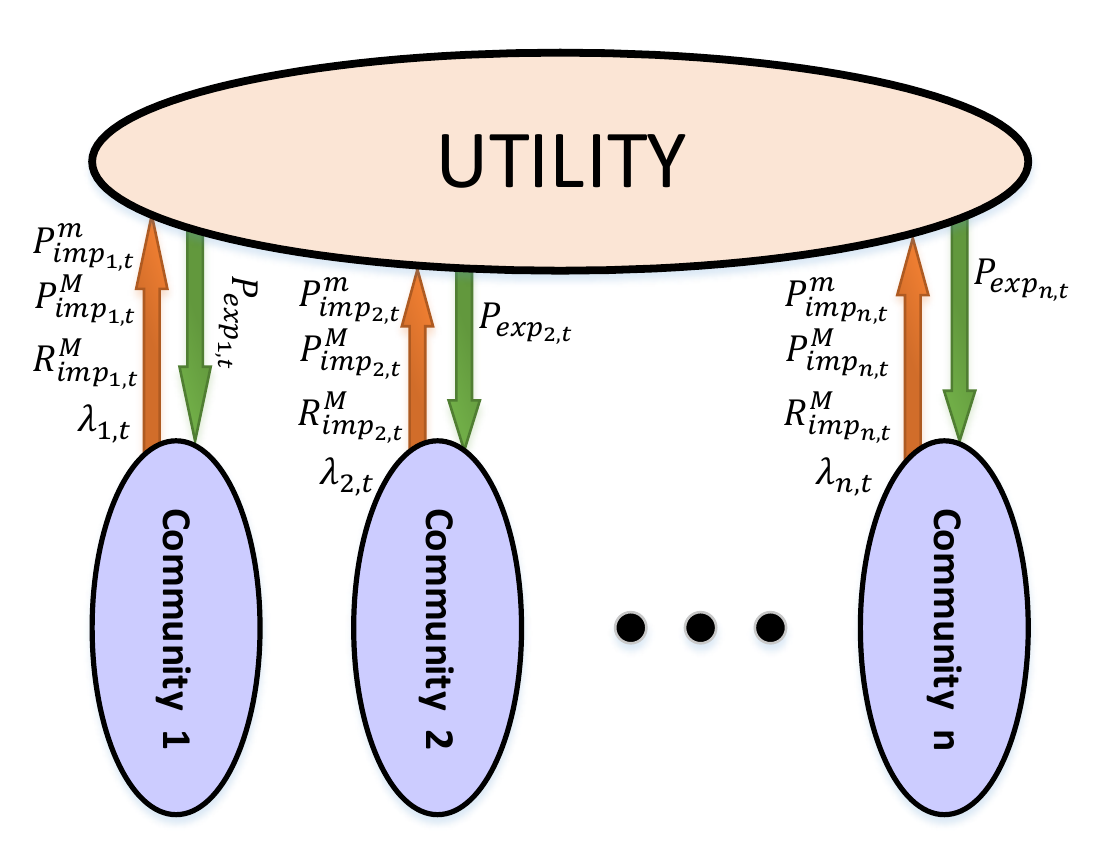}
	\caption{Information flow of multi-horizon LUBS algorithm. Information related to reactive power is not indicated in the figure.}
	\label{fig:LUBSinfoflow}
\end{figure}

In this structure, communities are responsible to determine the optimal price vectors at different hours to be fed into the utility's OPF problems. At each iteration, communities also update utility about the maximum and minimum power and reserve they can provide. The information flow structure based on LUBS algorithm is illustrated in Fig. \ref{fig:LUBSinfoflow}.

\subsubsection{Community Optimization and Price Updating}
Once the communities receive the utility's power import information for 24 hours, they must determine their local generator and battery dispatch four 24 hours and announce the corresponding price signals back to the utility.


In general, the community connected to the bus $i \in \mathcal{A}$ determines its Lagrange multipliers corresponding to the power balance constraints $(\lambda^{p}_{i,t},\lambda^{q}_{i,t})$. When the generator limits are not binding, the Lagrangian multipliers or the prices are same as the marginal prices of the generators.

One of the requirements in LUBS method is that the solution sought by the utility for power and reserve must be feasible for communities. If Community $i$'s marginal cost is cheaper than the minimum marginal cost of the utility and other communities, the utility will demand its maximum generation and battery power at all hours. This will be infeasible for the battery due to the energy constraints of the battery. Communities are then required to update the maximum and minimum power limits at each iteration in order to prevent such issues.

%

Community $i$ also updates the maximum reserve it can provide at each iteration according to its dispatched battery power as below and sends it to utility,
\begin{align}
  R_{{i,t}}^M(k+1)=  R^M_{g_i} - P^m_{b_{i}}+ P_{b_{i,t}}(k) \nonumber
\end{align}
\begin{figure*}[htbp]
	\centering
	\includegraphics[width=1\textwidth]{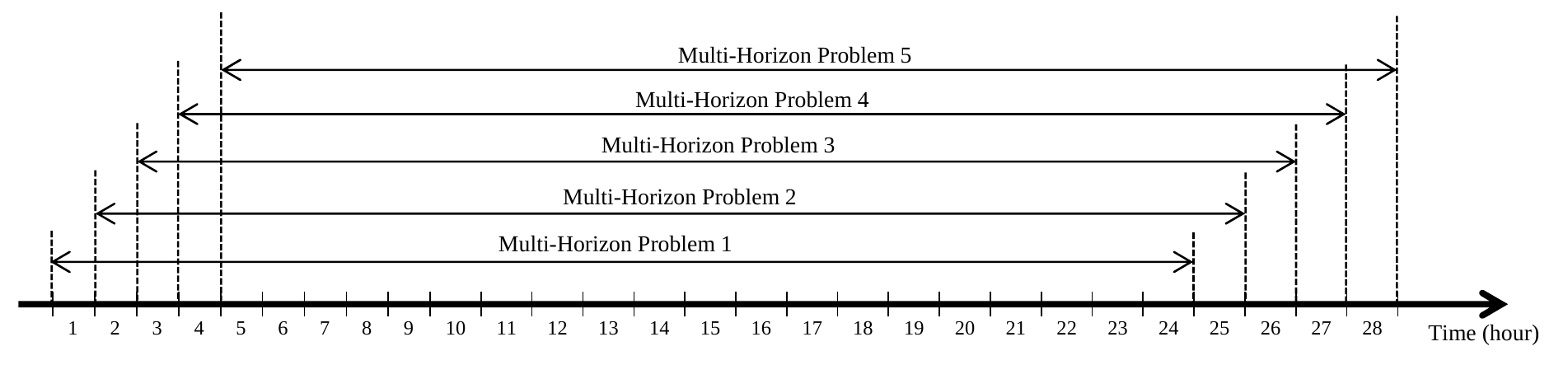}
	\caption{Time schedule for moving-horizon problems solving.}
	\label{moving-horizon-intro}
\end{figure*}
\subsubsection{Utility Optimization}
At iteration $k$, the utility solves its own subproblem given price values $\lambda_{i,t}(k)$ as well as power limits to determine the power import vector $P_{imp}$ for 24 hours while considering maximum and minimum limits of power and reserve which are announced by communities. The power import information is then announced to the communities. The procedure to solve utility OPF problem is the same as that in the subgradient method. The utility also computes the utility cost $C_u (k)$ at $k$-th iteration.

\subsection{Implementation and Moving Horizon Optimization}
 We assume a 24-hour PV profile and a 24-hour load profile based on forecast.  For the current hour, the load and PV outputs are known and the future 24 hours are predicted. The decision making procedures described above are then carried out for 24-hour iteratively. For real-time operation implementation, each iteration will be carried out for a time period, say two minutes. Every two minutes, the utility and the communities exchange information and update their decision variables. If an algorithm can converge in ten steps, the optimal operating condition can be realized in 20 minutes. This ``learning process'' is achieved by continuous information exchange and updating.

Suppose at Hour $k$, the learning takes 20 minutes and for the rest of the 40 minutes, both the communities and the utilities are keeping the process however the generator power output and the battery power output will be constant. Moving to Hour $k+1$, the load profile now has a change. The communities and the utility need to find the optimal battery dispatch and generator dispatch again through ``learning.''

The load profile and the PV profile at Hour $k+1$ at real-time are different from the prediction made at Hour $k$. Therefore, the load and PV profiles should be adjusted to reflect the $k+1$ hour information. Prediction for the future 24 hours needs to be made again. Another 24-hour optimization problem has to be solved iteratively.

The above mentioned procedure is termed as moving horizon optimization. To take the underlying power system dynamics into consideration, the power command decided by the decision making layer should better be continuous. Therefore, instead of assuming a same initial price for all time horizons (e.g, $\$0/MWh$), it is more desirable to start the next horizon ``learning'' process with the information from the previous time horizon.

\section{Convergence Analysis}

In this section, convergence properties of the two iterative methods are examined. For simplicity, the analysis is limited to energy price and real power only.
\subsection{Subgradient Method}
 In this method, the price $\lambda^k$ is announced to both the utility and a community at iteration $k$, then they decide how much power they would like to exchange with each other. Regarding the power level of both sides, the price update sector updates the price for the next iteration. If the slopes of marginal cost functions corresponding to utility and microgrid are respectively $a_1>0$ and $a_2>0$. Then the import and export power can be expressed as:
\begin{align}
P_{\rm imp}^k=P_{\rm imp}^0-\frac{1}{a_1}\lambda^k \label{Pimp}\\
P_{\rm exp}^k=P_{\rm exp}^0+\frac{1}{a_2}\lambda^k \label{Pexp}.
\end{align}
\eqref{Pimp} also indicates the economic behavior of the utility. The cheaper the price, the more power will be purchased. On the other hand, the greater the selling price, the more power will be sent out by the community as shown in \eqref{Pexp}.

\begin{figure}[htbp]
	\centering
	\includegraphics[width=0.35\textwidth]{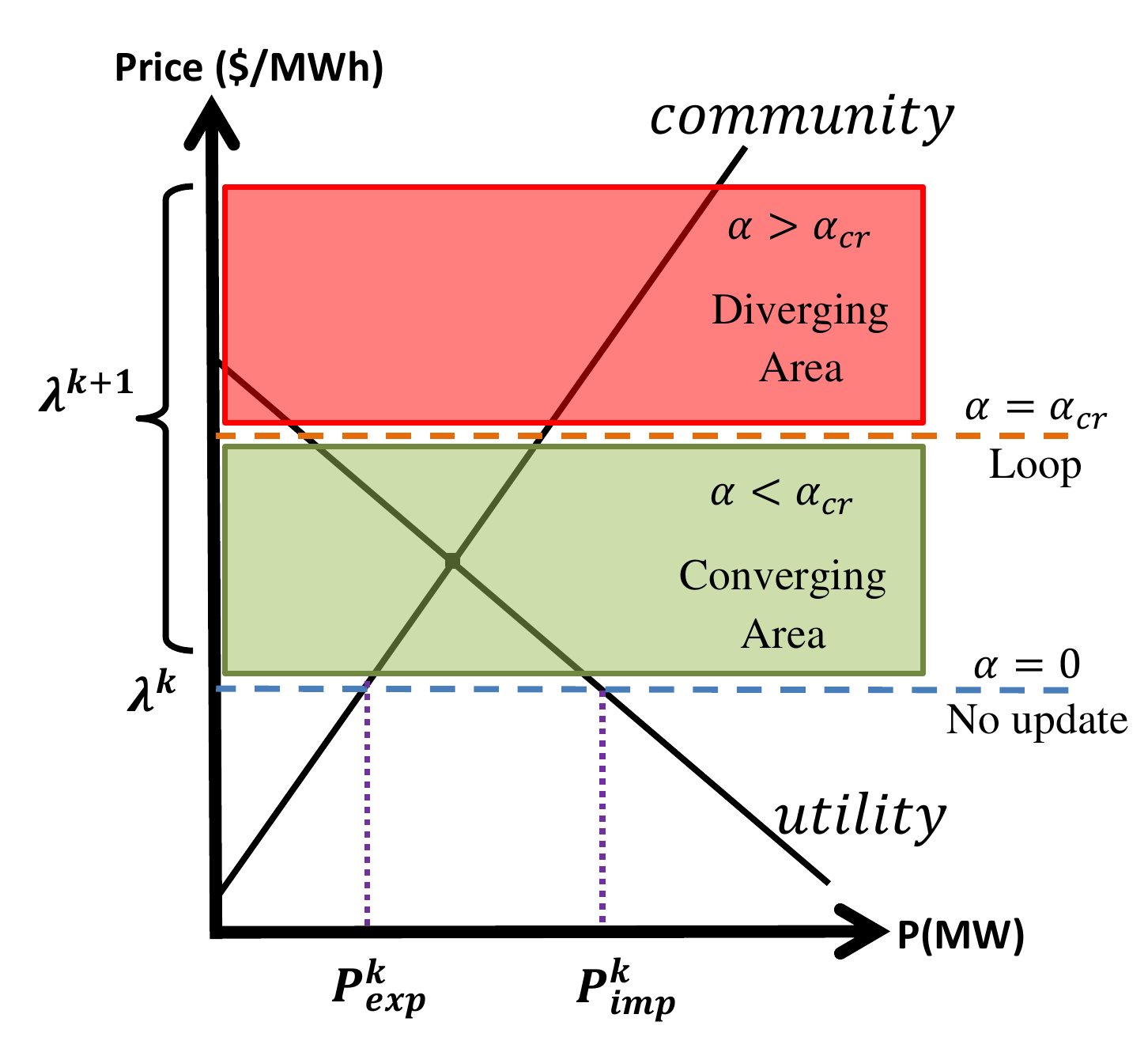}
	\caption{Graphical presentation of different regions of subgradient method convergence for various values of $\alpha$. The algorithm converges if $\alpha<\alpha_{cr}$, diverges if $\alpha>\alpha_{cr}$, and circulate in a loop if $\alpha=\alpha_{cr}$.}
	\label{SG_convergence}
\end{figure}

 Then we can obtain the iteration of the price.
 \begin{align}
\lambda^{k+1}=\left[ 1-\alpha \left(\frac{1}{a_1}+\frac{1}{a_2}\right)\right] \lambda^k + \alpha(P_{\rm imp}^0-P_{\rm exp}^0)
\end{align}
The iteration problems can be viewed as discrete domain dynamic problems. Hence, the convergence of the algorithm is the same as the stability of the corresponding discrete dynamic system. $\left|1-\alpha \left(\frac{1}{a_1}+\frac{1}{a_2}\right)\right|$ should be less than 1. Therefore, when $\alpha$ is very small, convergence is guaranteed. The critical value of $\alpha$ which is notated as $\alpha_{cr}=\frac{2a_1a_2}{a_1+a_2}$. When $\alpha<\alpha_{cr}$, subgradient method converges. When $\alpha>\alpha_{cr}$, the algorithm will not converge. For an $\alpha$ equal to critical value, the iterations will circulate in a loop. Fig. \ref{SG_convergence} illustrates the possible regions for $\lambda^{k+1}$ in a case where $P_{\rm imp}^k>P_{\rm exp}^k$ which is divided by the line $\lambda^{k+1}=\lambda^k+\alpha_{cr}(P_{\rm imp}^k-P_{\rm exp}^k)$ to converging and diverging areas.

\subsection{LUBS Method}

In this method, a community sets price based on the power demand by the utility. The utility then sets its import power based on the price.
Therefore, the iterative procedure can be explained as follows.
\begin{align}
P_{\rm imp}^k&=P_{\rm imp}^0-\frac{1}{a_1}\lambda^k \label{Pimp2}\\
P_{\rm imp}^k&=P_{\rm exp}^0+\frac{1}{a_2}\lambda^{k+1}\label{Pexp2}
\end{align}
\eqref{Pimp2} indicates that the utility sets its $k$-th step power based on a price $\lambda^k$. \eqref{Pexp2} indicates that the community updates its price based on the power demand from the utility.

\eqref{Pimp2} and \eqref{Pexp2} lead to the following iteration.
\begin{align}
\lambda_{k+1}=-\frac{a_2}{a_1}\lambda^k + a_2(P_{\rm imp}^0-P_{\rm exp}^0)
\label{eq:prediction}
\end{align}
The convergence criterion becomes $a_2<a_1$, or the slope of marginal cost of a community should be less than that of a utility. This condition is very restrictive. The algorithm is modified to achieve better convergence.

\begin{figure}[htbp]
	\centering
	\includegraphics[width=0.5\textwidth]{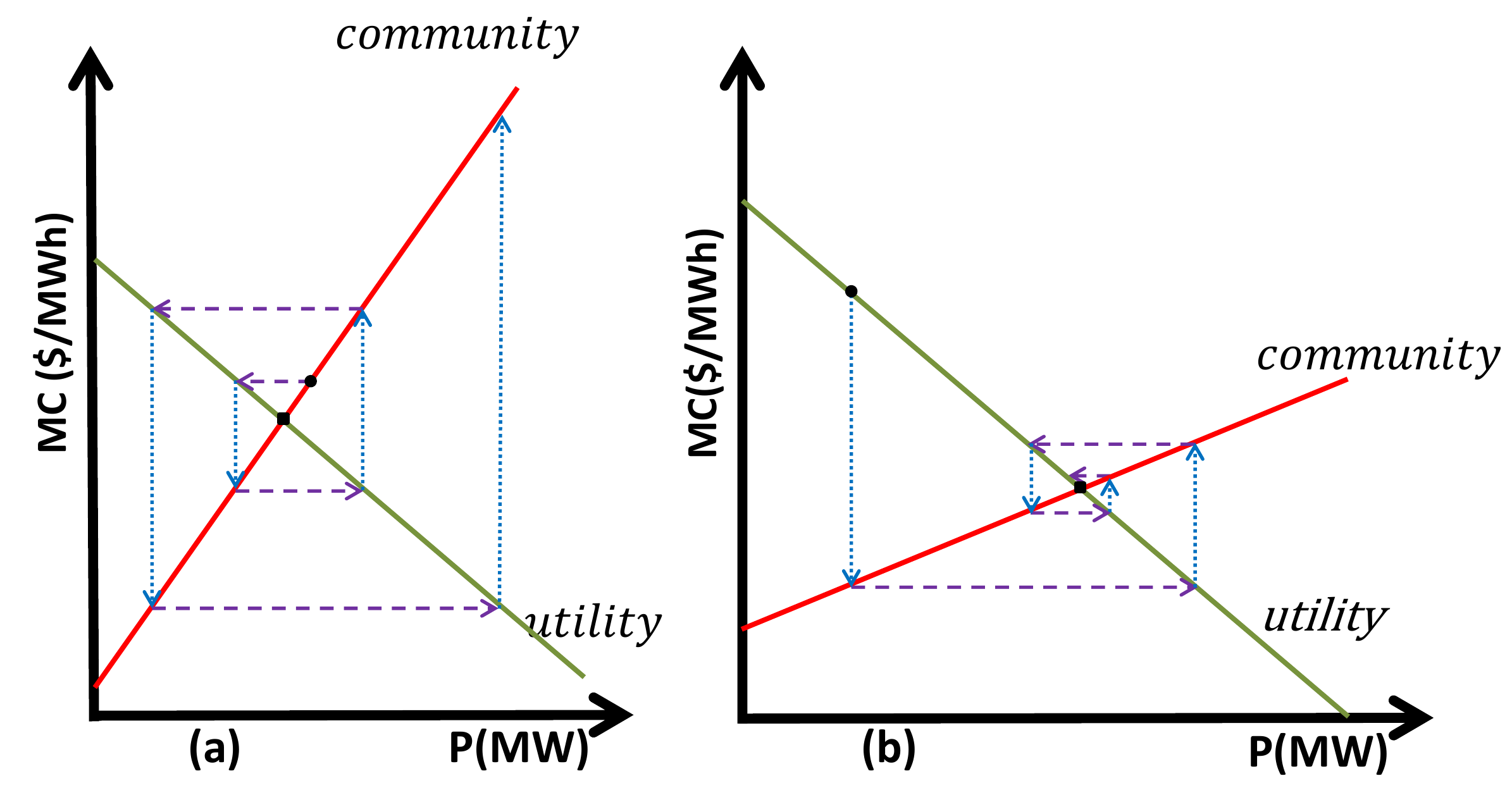}
	\caption{Graphical presentation of LUBS convergence for different cases, a) LUBS diverges when the slope of the utility's marginal cost is less than that of the community's, b) LUBS converges if the slope of the utility's marginal cost is greater than that of the community's.}
	\label{LUBS_convergence}
\end{figure}

 The scenarios are shown graphically in Fig. \ref{LUBS_convergence} where LUBS diverges when the community has more convex generation cost than microgrid does, \emph{i.e.},  $a_2>a_1$ (Fig \ref{LUBS_convergence}-a) and it converges otherwise when $a_2<a_1$ (Fig \ref{LUBS_convergence}-b). The case of $a_1=a_2$ is also considered as non-converging situation since the solutions will be trapped in a loop.

The price computed in \eqref{eq:prediction} is treated as a prediction ($\widetilde{\lambda}_{k+1}$). Based on this prediction, $100\sigma$ percentage of the update will be made. The $k+1$ step price will be:
\begin{align}
\lambda_{k+1} & = \lambda_{k+1} + \sigma(\widetilde{\lambda}_{k+1}-\lambda_k)\\
&=(1-\sigma)\lambda^k+\sigma\left(a_2(P_{\rm imp}^0-P_{\rm exp}^0)-\frac{a_2}{a_1}\lambda^k\right)\\
& =\left(1-\sigma \frac{a_1+a_2}{a_1}\right)\lambda^k+ \sigma a_2(P_{\rm imp}^0-P_{\rm exp}^0)
\end{align}
The convergence criteria is then identified as whether $\left| 1-\sigma\frac{a_1+a_2}{a_1}\right|<1$ or not. The critical value of $\sigma$ can be explicitly derived as $\sigma_{cr}=2\frac{a_1}{a_2+a_1}$. LUBS algorithm converges if $\sigma<\sigma_{cr}$, diverges when $\sigma>\sigma_{cr}$. Iteration will betrapped in a loop if $\sigma=\sigma_{cr}$.

\begin{figure}[htbp]
	\centering
	\includegraphics[width=0.35\textwidth]{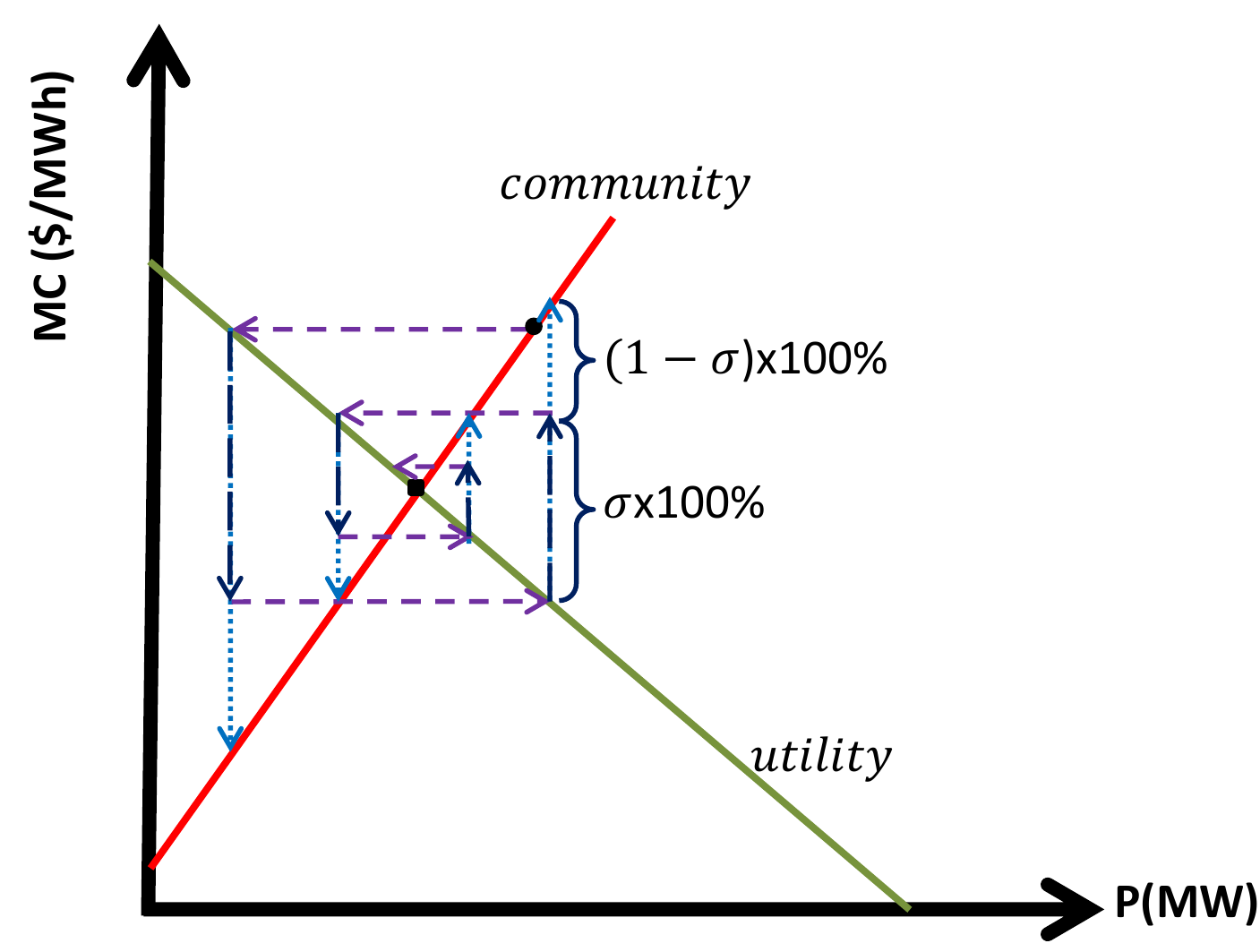}
	\caption{Graphical presentation of LUBS convergence for different cases, a) LUBS diverges when agent 1's generation cost is less convex than generator 2's, b) LUBS converges if agent 1's generation cost is more convex than generator 2's.}
	\label{LUBS_convergence_sigma}
\end{figure}

Fig. \ref{LUBS_convergence_sigma} shows $\sigma$ modification makes the algorithm converge to the optimal point for the case where $a_1<a_2$ (Fig \ref{LUBS_convergence}-a).

\section{Case Study}
The proposed two decision making paradigms have been implemented in a case study of a radial 42-bus test feeder introduced in IEEE Standard 399 \cite{IEEE399}. One community is connected on Bus 50. The generator cost functions, power and reserve limits are described in Table \ref{table_g}.
\begin{table}[htbp]
\centering
\caption{Parameters and cost functions of generators in IEEE Std. 399 test feeder $C(P_{g})=0.5\alpha P_{g}^2+\beta P_{g}+\gamma$}
\begin{tabular}{|c|c|c|c|c|c|c|c|}
  \hline
  Bus \# &Owner&$P_g^m$&$P_g^M$&$R_g^m$&$\alpha$&$\beta$&$\gamma$\\
  \hline
  100&Utility&0&2&0&0.1&55&0\\
  \hline
  4&Utility&0&12&2&0.3&50&0\\
  \hline
  9&$\mu G1$&0&5&0&0.4&42&0\\
  \hline
  29&$\mu G2$&0&3&0&0.4&35&0\\
  \hline
  30&$\mu G3$&0&4&0&0.2&38&0\\
  \hline
  50&$\mu G4$&0&11&8.8&0.2&49&0\\
  \hline
\end{tabular}
\label{table_g}
\end{table}

\begin{figure}[htbp]
	\centering
	\includegraphics[width=0.5\textwidth]{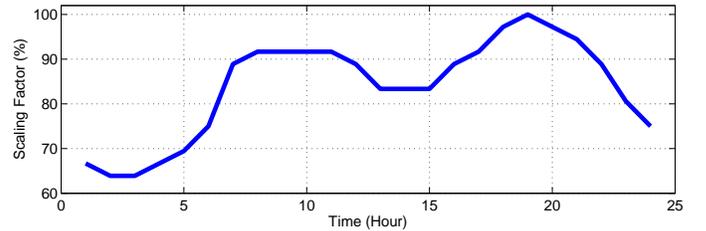}
	\caption{Demand scaling factor in 24-hour horizon, applied to obtain power demand profile.}
	\label{LoadFactor}
\end{figure}

The temporal demand profile of the entire test feeder is obtained by multiplied by a time-variant scaling factor $\rho(t)$ as shown in Fig. \ref{LoadFactor}.

The utility optimization problem for 24 hours is solved in MATPOWER for 24 snapshots. The community optimization problem is carried out by CVX \cite{cvx}. Simulation results are presented in the following figures Figs. \ref{sim_price}-\ref{sim_reserve}. The 24-hour profiles or 8-hour profiles clearly demonstrate how the ``learning'' processes took place.

\begin{figure}[htbp]
	\centering
	\includegraphics[width=3.5in]{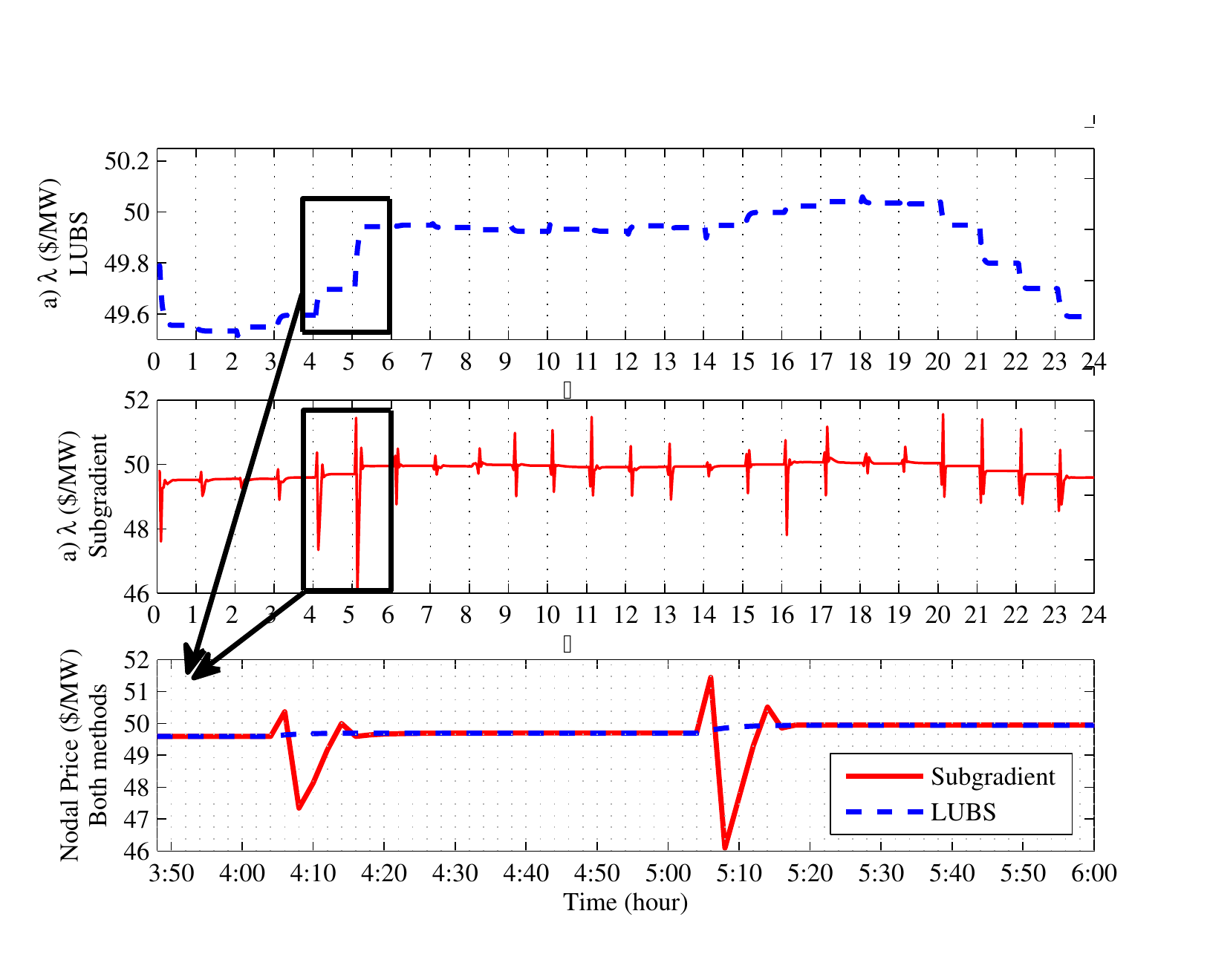}
	\caption{Comparison of community nodal price behavior through iterations in LUBS and subgradient for 24 consecutive moving-horizon simulations, where each iteration is assumed to take two minutes. a) The price in LUBS method throughout an entire day, b) the price in subgradient method throughout an entire day, c) a closer look on the price in both methods during a
two-hour horizon.}
	\label{sim_price}
\end{figure}
\begin{figure}[htbp]
	\centering
	\includegraphics[width=3.5in]{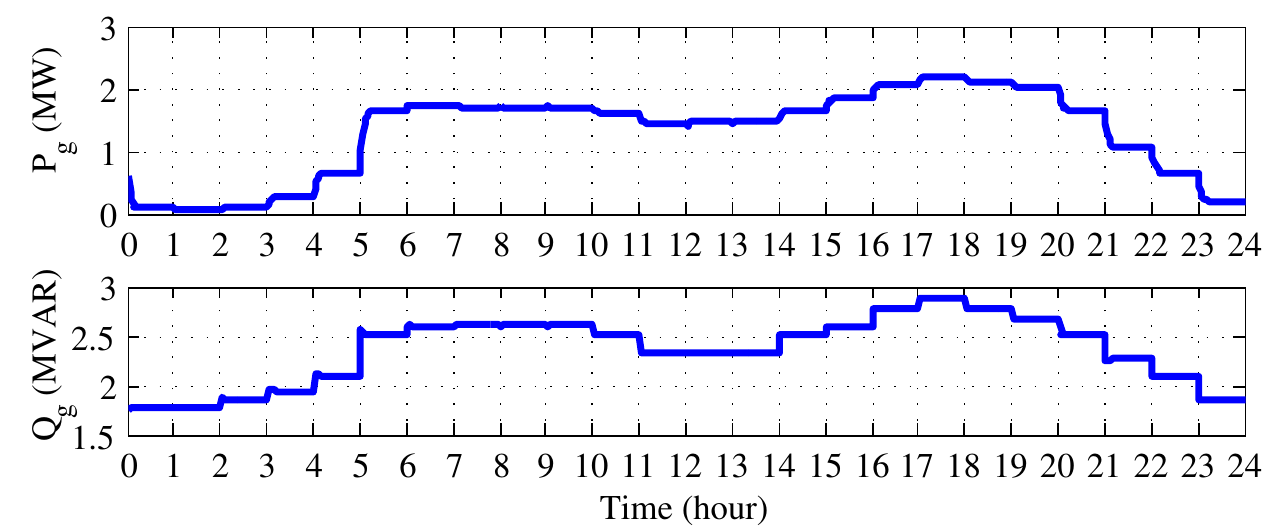}
	\caption{A utility generator's power output profile.}
	\label{sim_utility_power}
\end{figure}
\begin{figure}[htbp]
	\centering
	\includegraphics[width=3.5in]{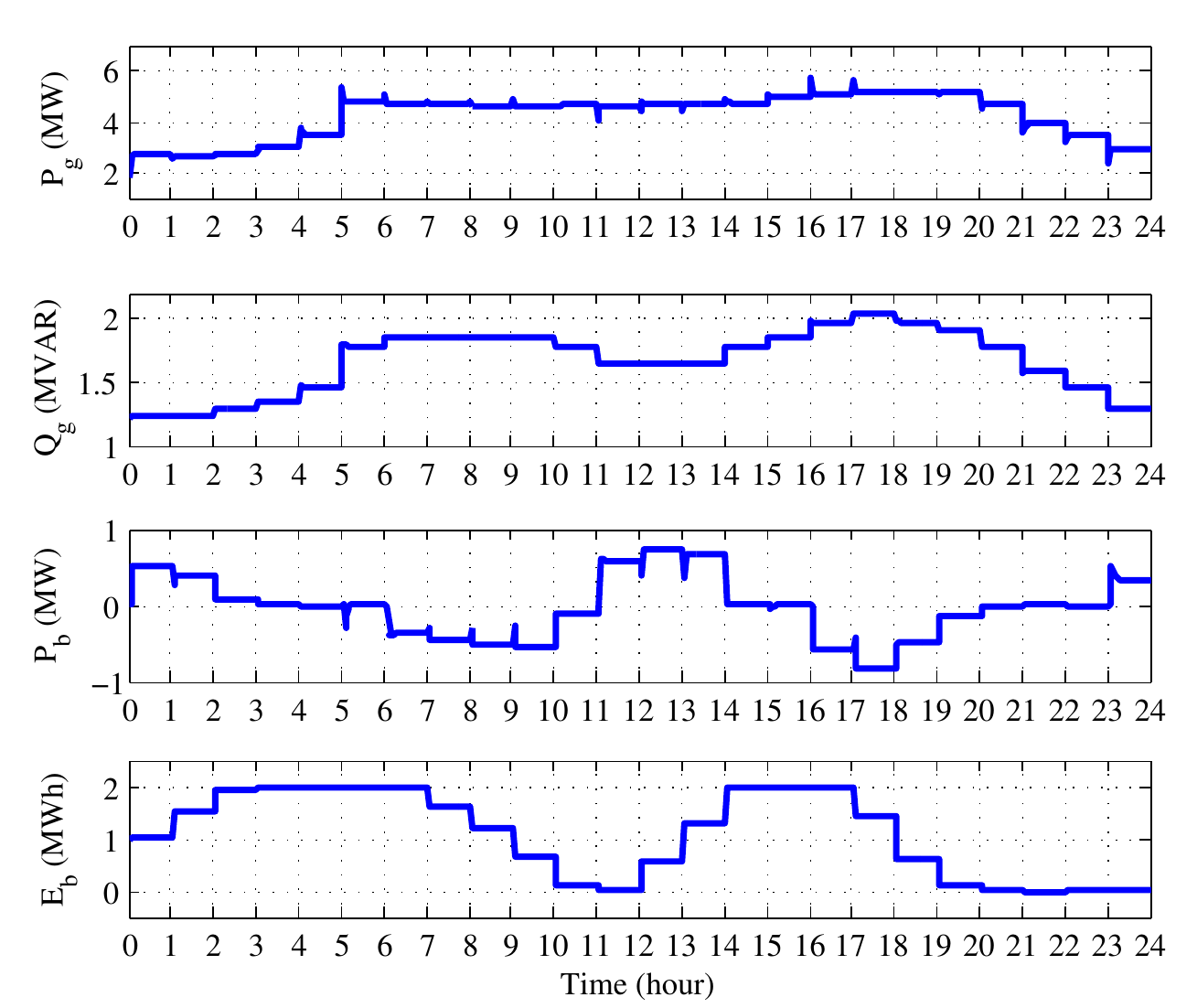}
	\caption{The community's generator and battery profile.}
	\label{sim_community}
\end{figure}

\begin{figure}[htbp]
	\centering
	\includegraphics[width=3.5in]{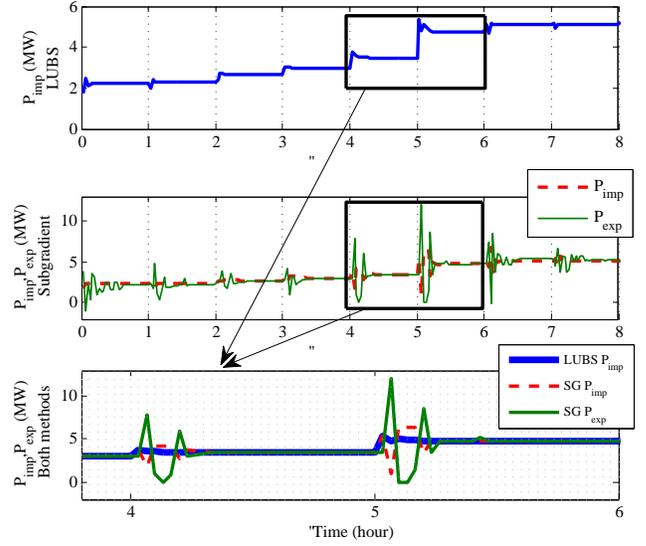}
	\caption{8-hour importing/exporting power profiles.}
	\label{sim_Pimpexp}
\end{figure}

\begin{figure}[htbp]
	\centering
	\includegraphics[width=3.5in]{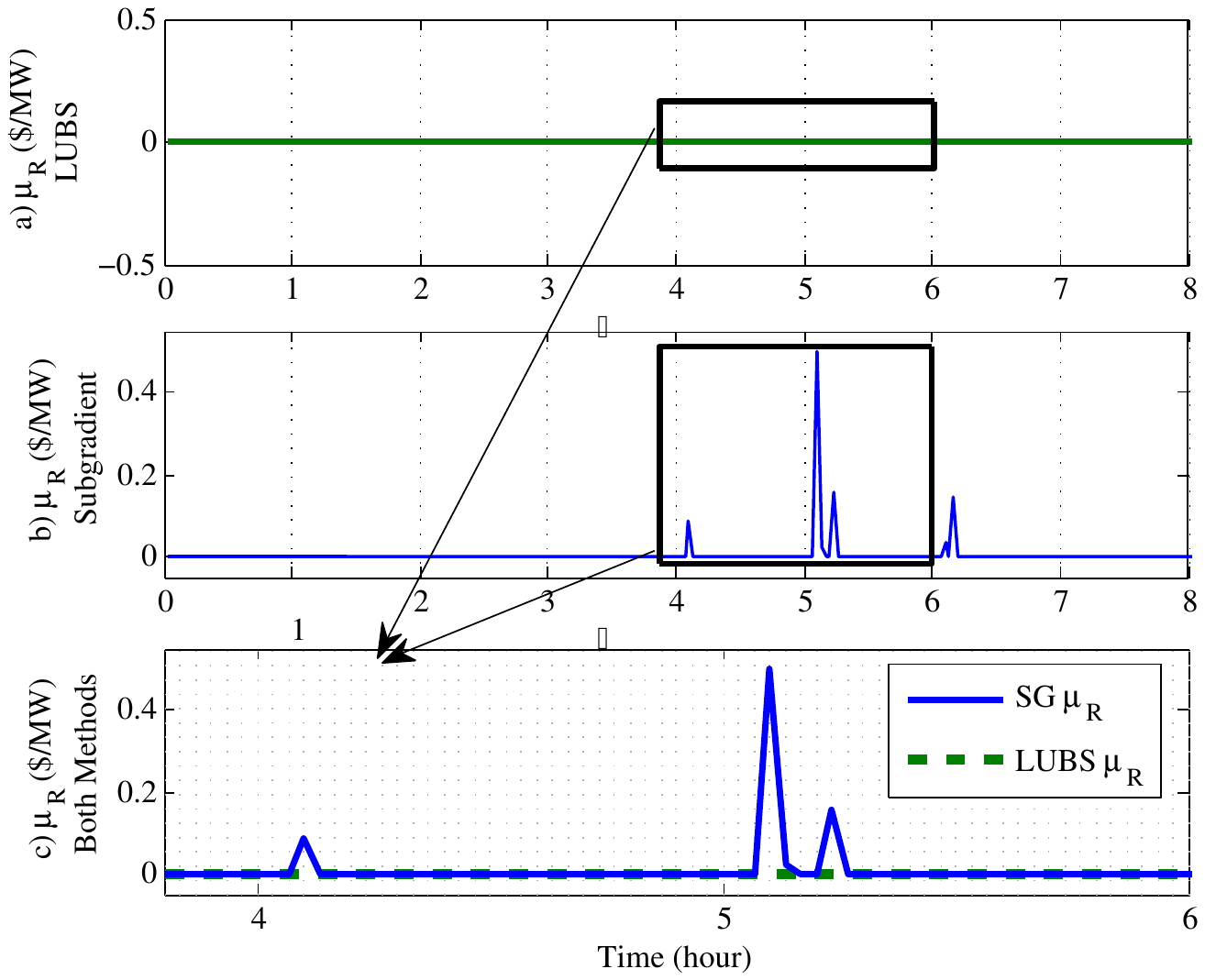}
	\caption{8-hour reserve price profile.}
	\label{sim_mu}
\end{figure}

\begin{figure}[htbp]
	\centering
	\includegraphics[width=3.5in]{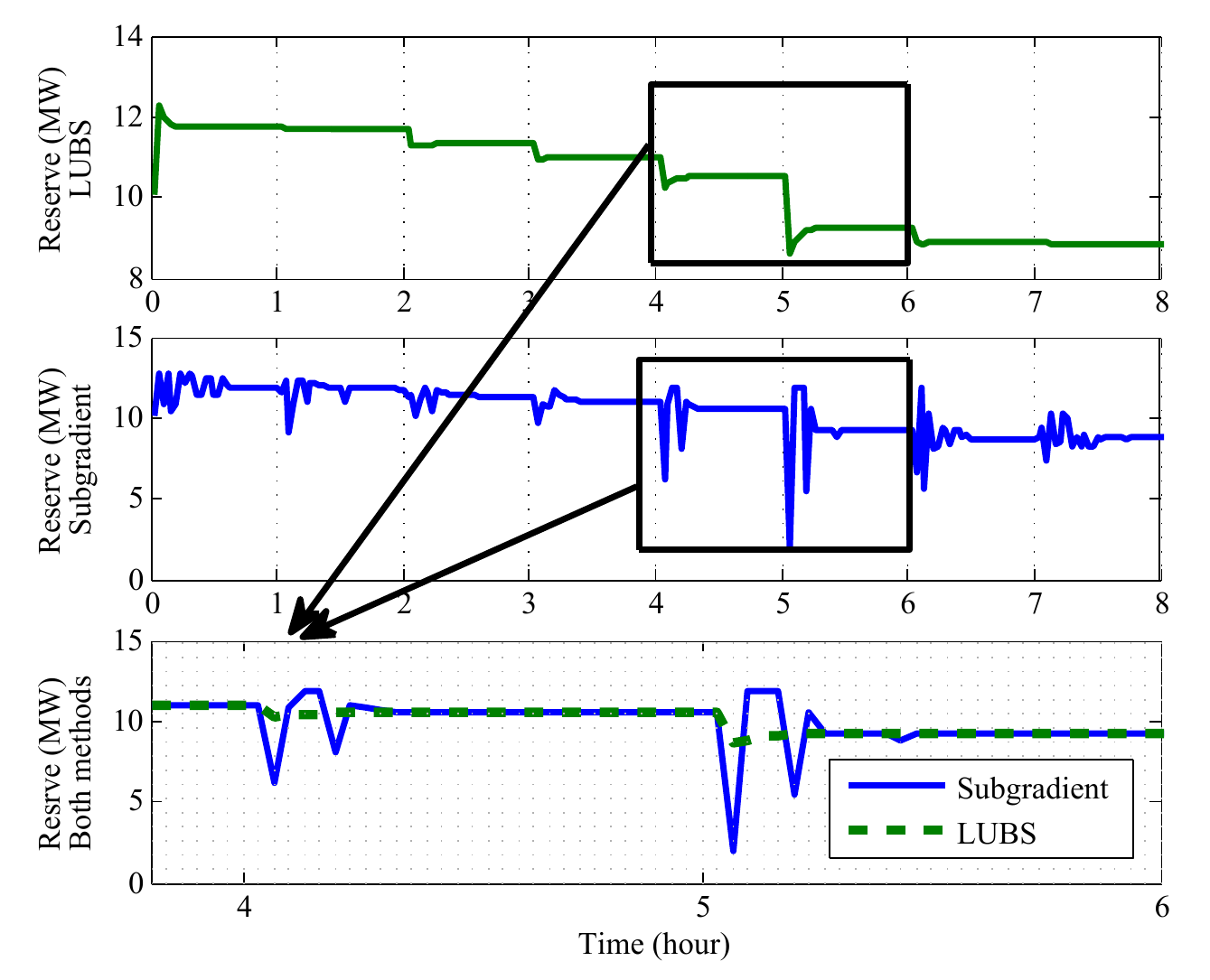}
	\caption{8-hour reserve profile.}
	\label{sim_reserve}
\end{figure}
Remarks: the comparison of LUBS and subgradient-based method shows that for real-world implementation, LUBS is more favorable due to the smoother change in the price signals and power levels.

\section{Conclusion}
Dual decomposition-based privacy preserving decision making paradigms for utility-community interaction for multiple horizons are developed. The developed paradigms enable utility and communities to make their own decisions based on limited information exchange. The paradigms are suitable for radial connected utility and communities. Implementation for real-time operation scenarios takes care of forecast and real-time discrepancy by using moving horizon optimization technique and by initiating the next horizon variables based on the information from the previous horizon. The information flow architectures are explained and the convergence properties are investigated. The proposed decision making architectures were tested by a case study and demonstrate the ``learning'' processes taken by a utility and a community.

\bibliographystyle{IEEEtran}
\bibliography{IEEEabrv,Decomposition}

\begin{IEEEbiographynophoto}{Vahid R. Disfani}
received his B.S. degree from Amirkabir University of Technology (Tehran, Iran) in 2006, and his M.S. degree from Sharif University of Technology (Tehran, Iran) in 2008.
Currently, he is pursuing a Ph.D. degree at University of South Florida, Tampa, FL. His research interests include Smart Grid, integration of renewable energy resources to microgrids, and electricity markets.
\end{IEEEbiographynophoto}

\begin{IEEEbiographynophoto}{Zhixin Miao} (S'00 M'03 SM'09) received the
B.S.E.E. degree from the Huazhong University of
Science and Technology,Wuhan, China, in 1992, the
M.S.E.E. degree from the Graduate School, Nanjing
Automation Research Institute, Nanjing, China, in
1997, and the Ph.D. degree in electrical engineering
from West Virginia University, Morgantown, in
2002.

Currently, he is with the University of South
Florida (USF), Tampa. Prior to joining USF in 2009,
he was with the Transmission Asset Management
Department with Midwest ISO, St. Paul, MN, from 2002 to 2009. His research
interests include power system stability, microgrid, and renewable energy.
\end{IEEEbiographynophoto}

\begin{IEEEbiographynophoto}{Lingling Fan}
received the B.S. and M.S. degrees in electrical engineering from Southeast University,
Nanjing, China, in 1994 and 1997, respectively, and the Ph.D. degree in electrical engineering
from West Virginia University, Morgantown, in 2001.
Currently, she is an Associate Professor with the University of South Florida, Tampa, where she has
been since 2009. She was a Senior Engineer in the Transmission Asset Management Department, Midwest
ISO, St. Paul, MN, form 2001 to 2007, and an Assistant Professor with North Dakota State University,
Fargo, from 2007 to 2009. Her research interests include power systems and power electronics. Dr. Fan serves as a technical program committee chair for IEEE Power System Dynamic Performance Committee and an editor for IEEE Trans. Sustainable Energy.
\end{IEEEbiographynophoto}
\begin{IEEEbiographynophoto}{Bo Zeng}(M'11) received the Ph.D. degree in industrial engineering from
Purdue University, West Lafayette, IN, with emphasis on operations research
He is an Assistant Professor in the Department of Industrial and Management
Systems Engineering at the University of South Florida, Tampa. His research interests
are in the polyhedral study and computational algorithms for stochastic
and robust mixed integer programs, coupled with applications in energy, logistics,
and healthcare systems.
Dr. Zeng is a member of IIE and INFORMS.
\end{IEEEbiographynophoto}

\end{document}